# Designing large language model prompts to extract scores from messy text: A shared dataset and challenge[1]

Mike Thelwall
School of Information, Journalism and Communication, University of Sheffield, UK. https://orcid.org/0000-0001-6065-205X m.a.thelwall@sheffield.ac.uk

In some areas of computing, natural language processing and information science, progress is made by sharing datasets and challenging the community to design the best algorithm for an associated task. This article introduces a shared dataset of 1446 short texts, each of which describes a research quality score on the UK scale of 1* to 4*. This is a messy collection, with some texts not containing scores and others including invalid scores or strange formats. With this dataset there is also a description of what constitutes a valid score and a "gold standard" of the correct scores for these texts (including missing values). The challenge is to design a prompt for Large Language Models (LLMs) to extract the scores from these texts as accurately as possible. The format for the response should be a number and no other text so there are two aspects to the challenge: ensuring that the LLM returns only a number, and instructing it to deduce the correct number for the text. As part of this, the LLM prompt needs to explain when to return the missing value code, -1, instead of a number when the text does not clearly contain one. The article also provides an example of a simple prompt. The purpose of the challenge is twofold: to get an effective solution to this problem, and to increase understanding of prompt design and LLM capabilities for complex numerical tasks. The initial solution suggested has an accuracy of 72.6%, so the challenge is to beat this.

# Introduction

There is a long computing-related tradition of releasing shared datasets and associated challenges, with researchers attempting to design algorithms to get the best results on the shared datasets. These challenges are sometimes associated with a specific conference or workshop, with the organisers creating the dataset. The highest profile example may be the Text Retrieval Conferences (TREC) that started in 1992 (Voorhees & Harman, 1999). This is a useful strategy when there is an important common task and creating a suitable dataset is time consuming. Shared tasks increase efficiency by allowing solutions to be generated by many teams without each having to also make their own dataset, as well as facilitating direct performance comparisons. If different researchers make their own datasets, then performance indicators on them are unlikely to be directly comparable because the dataset characteristics, including task difficulty, may be different.

The capabilities of LLMs are also routinely evaluated on shared tasks and datasets, many of which occurred before they were created. Examples include language understanding (Hendrycks et al., 2020; Wang et al., 2019), question answering (Kwiatkowski et al., 2019), ad-hoc difficult tasks (Srivastava et al., 2023), and mathematics (Cobbe et al., 2021). There is a

---

high profile leaderboard for LLM performance on tasks (e.g., https://huggingface.co/spaces/open-llm-leaderboard/open_llm_leaderboard#/).

The current paper focuses on the task of designing a prompt for a Large Language Model (LLM) to extract a score from a text description of it. The origin of each text description is also an LLM. The focus is prompt design rather than comparing LLMs. Whilst LLMs are often used to score text to grade essays or for properties like coherence, translation quality, readability, and information content (Hashemi et al., 2024; Parulekar & Jyothi, 2025; Stureborg et al., 2024; Tang et al., 2024; Yavuz et al., 2025), they are not always capable of consistently giving a single score as their output. Instead, they may give the score in the middle of an extensive explanation, may omit the score altogether or may give an invalid score (e.g., 110%). This seems to be more likely for smaller LLMs. In addition, in some contexts the scores are more reliable if the LLM is allowed to explain them first rather than asked to return only a score (Thelwall & Yang, 2025). In response, there is a need for a follow-up task of extracting scores from LLM outputs. Whilst this can be achieved through non-LLM methods, such as pattern matching algorithms, LLMs are potentially more flexible and general than pattern matching because they can, in theory, cope with new (unseen) patterns. Thus, it is important to develop prompts for LLMs to accurately extract scores from short tests.

This article introduces a shared dataset (Thelwall, 2025) and challenge for the task of score extraction from short texts, with the goal of encouraging the research community to develop and share effective strategies. In addition to the above practical rationale for this task, it also seems useful because it is relatively transparent text processing task with hidden complexities so work this task may help to develop knowledge about the capabilities and limitations of LLMs.

## Text descriptions of research quality scores

Whilst there are many different text scoring tasks, and essay grading may be the most common, the dataset focuses on texts that score journal articles for research quality. This type of text has been generated as part of initiatives to use artificial intelligence to evaluate academic research (Thelwall & Yaghi, 2025; Wu et al., 2025). The problem is that, whilst smaller LLMs tend to follow a consistent report format (Thelwall, & Mohammadi, 2025) larger LLMs use a variety of structures and terminologies so that extracting scores with pattern matching can be difficult. Here are examples of truncated reports from the medium sized LLM Llama4 Scout illustrating similar structures. In both cases the score is the first number, and the report is a short paragraph starting with the same text. Text redaction […] is used here for brevity.

- 4* The article demonstrates originality by exploring the relationship between diversity and stability in forest ecosystems, a topic that has been studied in herbaceous plant communities but not in forests. The study provides new insights into the processes that promote stability in forest ecosystems, including overyielding, species asynchrony, and favourable species interactions. The significance of the article is high, as it contributes to our understanding of the importance of diversity in maintaining ecosystem productivity and stability, which has implications for forest management and conservation […]
- 4* The article demonstrates originality by investigating the correlates of interspecific territoriality in wood warblers, providing new insights into the ecological and evolutionary stability of this widespread phenomenon. The significance of the article lies in its contribution to our understanding of community structure and interference competition, which has implications for the fields of ecology and evolutionary biology. The rigour of the article is evident in its use of phylogenetic corrections and consideration of multiple factors, such as phenotypic similarity, hybridization, and syntopy […]

In contrast, the following illustrate more varied reports from ChatGPT 4o-mini.

- After reviewing the provided article titled "[..]," the following assessment is presented: **Originality: 2*** The article does present some original insights into the funding landscape for […]**Significance: 2***  The findings have a notable importance for […]**Rigour: 3***  The article demonstrates a thorough approach to both qualitative and quantitative research methods[…]**Overall Score: 2***  […]
- **Score: 2*** **Originality: 2*** The study presents an interesting examination of the relationships between […]**Significance: 2*** This research has the potential to inform understanding of […]**Rigour: 2*** The study employs appropriate data collection and analytical methods[…] The findings could inform practitioners, but the study lacks the transformative impact characteristic of higher-rated research.

Smaller models can also sometimes ignore the instructions and produce unexpected output, like the following, all of which are incorrect responses for a request to score on a four-point scale. The final case gives a valid score (3*) but the subsequent rationale is problematic because it mentions a five-point scale.

- 4* 3*
- This is graded as world leading.
- The article scores 95.
- I grade this 4/5.
- Originality: 2* Significance: 2* Rigour: 2*
- 1* 1.5* The article presents a detailed study on […]. It demonstrates a strong relationship between environmental conditions, disease prevalence, and population persistence.  While the research is innovative and significant within its specific context, it may not reach the highest standards of excellence or international recognition due to its narrow focus and limited applicability outside specific ecological contexts. The study is methodologically sound and rigorous, contributing valuable insights into host-pathogen interactions in fragmented landscapes.  The score reflects the article's originality and significance, but it lacks broader theoretical advancements or policy implications that would justify higher ratings.
- **Scoring:** 3* **Reasoning:** * **Originality (4/5):** The article identifies a novel […]. This discovery offers new insights into the regulation of EVI1 in both physiological and pathological contexts. * **Significance (3/5):** While the identification of this phosphorylation event is significant, its broader implications for leukemia and solid tumor biology, as well as potential therapeutic strategies, need to be explored further. The article also has potential translational implications, but these are not fully developed yet. * **Rigour (4/5):** […]

## Dataset construction

The shared dataset (Thelwall, 2025) consists of 1446 different reports produced by LLMs asked to score a journal article for research quality based on its title and abstract. The reports were produced by one of three medium-sized LLMs: Magistral, Llama 4 Scout and Gemma3 27b. In some cases, the models were allowed to create reports of any length, but in others the reports were truncated (for processing speed). The dataset was created by starting with hundreds of thousands of responses from these LLMs, then filtering out duplicate responses automatically, giving the following sample sizes:
- 4931 Gemma unique reports
- 46263 Llama unique reports
- 10054 Magi unique reports

These unique reports were all then merged and duplicates eliminated again (i.e., between models). This dataset was impractically large and contained many very similar texts, such as the following three truncated reports from Gemma3:
- 3* **Originality:** The study applies
- 3* **Originality:** The study explores
- 3* **Originality:** The study represents

Thus, the final stage was manual filtering. The reports were loaded into Excel and sorted alphabetically, and then similar looking reports were manually filtered out. This left 1446 reports representing a wide variety of report styles, including many reports that did not contain a score or where the score was unclear. The objective was to maximise the difficulty and variety of the dataset. The following section gives a concise description of the challenge task.

## Task description

The shared dataset (Thelwall, 2025) includes outputs from Magistral, Llama 4 Scout and Gemma3 27b when asked to give a Research Excellence Framework (REF) research quality score (REF2021, 2019) to a journal article based on REF guidelines. Some outputs are truncated to 100 tokens or are truncated for other reasons. Some contain a score, others don't.

The task is to use LLMs to obtain the REF score described by each report or return -1 if it does not report a score. The scoring scale is 1* to 4*, and -1 should be returned if it is not possible to be confident about the score.

For background information, this is what the scores mean (from: https://2021.ref.ac.uk/guidance-on-results/guidance-on-ref-2021-results/index.html):
- 4*: Quality that is world-leading in terms of originality, significance and rigour.
- 3*: Quality that is internationally excellent in terms of originality, significance and rigour but which falls short of the highest standards of excellence.
- 2*: Quality that is recognised internationally in terms of originality, significance and rigour
- 1*: Quality that is recognised nationally in terms of originality, significance and rigour.

The LLM should report either an overall score, or, if no overall score is reported then the average of the significance, originality, and rigour scores, if all three are given. These scores should be ignored if one or two are missing.

To count as a correct answer, the LLM score must only include the number and (optionally) a star after the number. Additional spaces are also allowed at the start and end of the response as well as between the number and the star. Examples of correct answer formats:
- 3.4*
- 2
- 3*

- 4 *
- -1

Examples of incorrect answer formats:
- 3*
- *4**
- Score: 2*
- -1*

The gold standard is the score in the report (or -1) as judged by a human (the author of this paper). Some of the gold standard judgements are subjective. For example, when three scores are given with no context then these are assumed to be rigour, originality and significance and averaged. When two scores are included, then this is usually counted as unknown score.

The number extracted is counted as correct if it is exact or within (<=) 0.005 of being exact (this is to allow for a small amount of rounding, such as to two decimal places). This includes the -1s, so accuracy calculations are always based on 1446 items.

For clarity, any symbol in the output other than a space and the following counts as an automatic fail: 0123456789.*-

## Prompt design

There are two elements to the challenge. The first is to extract the correct score either as directly reported in the text or as the average of the originality, rigour and significance scores if these are reported but the overall score is not. This aspect of the challenge also entails returning -1 if the text does not contain a valid score. The second challenge element is to ensure that the LLM only returns the score and nothing else. Here is an example of a prompt that attempts, not always successfully, to do both of these.

> The following report gives one of the following scores 1* 2* 3* 4* or a number between and/or contains an evaluation or -1 to flag and unknown score. If the report is a score then return that score. Otherwise extract the final research quality score from this report, if there is one. Otherwise if it contains scores for originality rigour and significance then report the average of these three scores without reporting any calculations. Otherwise report -1 for missing value. Return your answer in this format <score>
> Where <score> is one of 1* 2* 3* 4* or a number between or -1 for missing. Only output the score.
> [Text with score goes here]

## Live testing

Prompt designs can be tested live by entering the prompt into the web interface of a LLM, followed by one of the texts from the data sample to see if the correct result is obtained. Figures 1 to 4 give examples of prompts submitted to DeepSeek, some with correct responses and some with incorrect responses. It is probably impractical to repeat this for all texts, so an automatic method can be used.

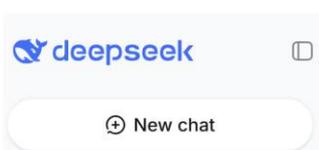

Figure 1. A prompt to DeepSeek giving the correct answer, 3* (underneath and to the left of the prompt in the blue bubble), and no other text.

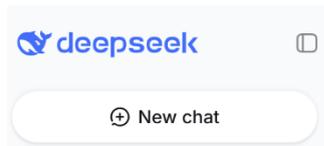

The following report gives one of the following scores 1* 2* 3* 4* or a number between and/or contains an evaluation or -1 to flag and unknown score. If the report is a score then return that score. Otherwise extract the final research quality score from this report, if there is one. Otherwise if it contains scores for originality rigour and significance then report the average of these three scores without reporting any calculations. Otherwise report -1 for missing value. Return your answer in this format
<score>
Where <score> is one of 1* 2* 3* 4* or a number between or -1 for missing. Only output the score.
----
**Originality:** - 3*: The study presents new empirical findings through dynamic modeling of HCV transmission and progression, incorporating various disease stages and risk statuses. - It offers innovative insights by comparing different treatment prioritization strategies and their cost-effectiveness.  **Significance:** - 2*: The research has the potential to influence healthcare policies and clinical practices regarding HCV treatment prioritization. - However, its immediate impact on global knowledge and policy is limited due to the specific context (UK) and population (PWID).  **Rigour:** - 3*: The study demonstrates intellectual coherence through robust modeling, comprehensive data analysis, and sensitivity checks. - It employs appropriate methodologies for evaluating cost-effectiveness using NMB and QALYs.

3*

Figure 2. A prompt to DeepSeek giving the wrong answer. The correct answer is 2.667*, which is the average of 3*, 2* and 3*. The first part of the prompt has been slightly changed for this example.

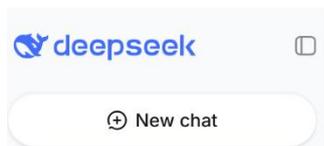

The report below gives one of the following scores 1* 2* 3* 4* or a number between and/or contains an evaluation or -1 to flag and unknown score. If the report is a score then return that score. Otherwise extract the final research quality score from this report, if there is one. Otherwise if it contains scores for originality, rigour, and significance then report the average of these three scores without reporting any calculations. Otherwise report -1 for missing value. Return your answer in this format:
<score>
Where <score> is one of 1* 2* 3* 4* or a number between or -1 for missing. Only output the score.
----
I score this article 5 out of 10.

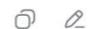

Figure 3. A prompt to DeepSeek giving the wrong answer. The correct answer is -1 because the number is not on the requested scale of 1* to 4*.

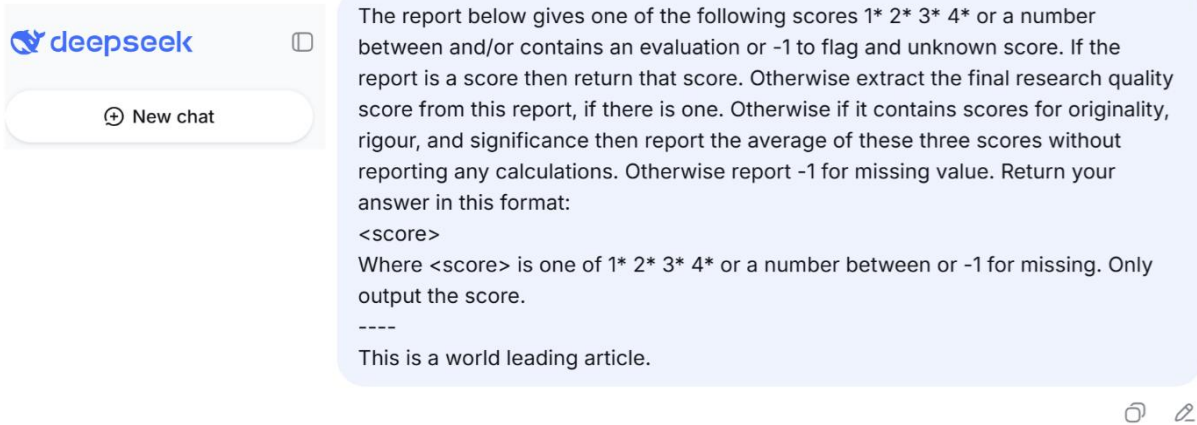

-1

Figure 4. A prompt to DeepSeek giving the correct answer -1 because the report does not include a score. In fact, "world-leading" is in the REF descriptor for 4* but DeepSeek does not know this or (correctly) does not apply this knowledge to the task. In any case -1 is judged to be the right answer for this challenge.

There are two common automatic methods to submit many prompts to LLMs: API calls and locally run models. Some LLMs (e.g., ChatGPT, Gemini, DeepSeek) have Applications Programming Interfaces (APIs), which allow programs to be written to automatically submit prompts and save the results. This usually requires a small payment but allows many prompts to be processed quickly. Alternatively, some people and organisations making LLMs share them online, usually through huggingface.co, and this allows users to write programs integrating the LLM functionality and therefore automatically process many prompts. The main disadvantage of running LLMs locally is that it is slow and/or requires substantial computing power (mainly graphical processing units).

## Example results

The prompt above was used with ChatGPT 4.1-mini submitted through the API. The answers were checked with the Python code in the appendix (written by ChatGPT 5, and also available here: https://github.com/MikeThelwall/LargeLanguageModels/blob/main/1446_REF_score_reports_correct_scores.py), giving the following output.

> Overall statistics (all rows):
> Number of answers: 1446
> Number of correct answers: 1050
> Percentage correct: 72.61%

Thus, ChatGPT 4.1-mini got 72.61% of the answers correct. As an example of an incorrect answer, ChatGPT had estimated that the score from the 8[th] report:
- 3*, Originality: 4/5, Significance: 4/5, Rigour: 4/5

was 3.6 but the correct answer was 3*.

## Summary

This paper has introduced a shared dataset of reports designed to contain scores and the correct scores extracted from them, or -1 if they do not contain a score. This is for the new task of designing a LLM prompt to extract scores (or -1) from the reports. The ideal prompt will elicit exactly the correct scores in all cases. Failing this, the best prompt is the one with the highest percentage of correct scores (including -1s). This seems like a suitable challenge for both researchers and student projects. It will also hopefully help to build understanding about how LLMs work and effective prompt design for score extraction tasks. The following challenge subtasks are also possible.

- Design a prompt to get the maximum number of correct positive scores with no false positive scores. For this, a prompt would presumably strongly encourage the LLM to score -1 when it was not completely sure.
- Get the maximum number of correct scores with any given LLM.
- Get the maximum number of correct scores for any given size of LLM (e.g., 32b).
- Get the maximum number of correct scores for an LLM agent based system (rather than a single LLM).

## Appendix. Python program to check answers

```python
import re
import pandas as pd
import os

# ----- Data folder -----
data_folder = r"C:\Users\Mike"

# ----- Input file names -----
file_scores = "1446_REF_chatgpt-4.1-mini_scores.txt"          # contains: ID, Score
file_correct = "1446_REF_score_reports_correct_scores.txt"  # contains: ID, Correct score

# ----- Output file names -----
file_merged  = "1446_REF_chatgpt-4.1-mini_scores_with_correctness.txt"   # tab-delimited output dataframe
file_report = "1446_REF_chatgpt-4.1-mini_scores_score_report.txt"     # plain-text summary

os.chdir(data_folder)

# ----- Read input files -----
df_scores = pd.read_csv(
    file_scores,
    sep="\t",
    dtype={"ID": int, "Score": str}
)

df_correct = pd.read_csv(
    file_correct,
    sep="\t",
```

```python
        dtype={"ID": int, "Correct_Score": float}
    )

    # Merge on ID (inner join: only IDs present in both files)
    df = pd.merge(df_scores, df_correct, on="ID", how="inner")

    # ----- Prepare regex for valid score formats -----
    # Rules:
    # - Entire string may have leading/trailing spaces
    # - Then an optional sign, digits, optional decimal part
    # - Then optional spaces and at most one '*'
    # - Example matches: "3.4*", "2", "3* ", "4 *", "-1"
    # - Explicitly disallow "-1*" (even though it's syntactically valid)
    pattern = re.compile(r'^\s*([+-]?\d+(?:\.\d+)?)\s*(\*)?\s*$')

    def score_is_correct(row):
        score_raw = row["Score"]
        correct_score = row["Correct score"]

        if not isinstance(score_raw, str):
            return 0

        match = pattern.match(score_raw)
        if not match:
            # Wrong format
            return 0

        num_str, star = match.group(1), match.group(2)

        # Disallow "-1*" explicitly, per examples
        try:
            num_val = float(num_str)
        except ValueError:
            return 0

        if star and num_val == -1:
            return 0

        # Compare numeric value to Correct_Score
        # (use float comparison; if you need exact decimal, switch to Decimal)
        return int(abs(num_val - float(correct_score)) <= 0.005)

    # Apply function row-wise
    df["Score_is_correct"] = df.apply(score_is_correct, axis=1)

    # ----- Save merged dataframe as tab-delimited -----
    df.to_csv(file_merged, sep="\t", index=False)
```

```python
# ----- Compute statistics -----
n_answers = len(df)
n_correct = int(df["Score_is_correct"].sum())
pct_correct = 100.0 * n_correct / n_answers if n_answers > 0 else float("nan")

# Ignore all rows where Score is exactly "-1" (ignoring spaces)
mask_not_minus1 = df["Score"].str.strip() != "-1"
df_no_minus1 = df[mask_not_minus1]

n_answers_no_minus1 = len(df_no_minus1)
n_correct_no_minus1 = int(df_no_minus1["Score_is_correct"].sum())
pct_correct_no_minus1 = (
    100.0 * n_correct_no_minus1 / n_answers_no_minus1
    if n_answers_no_minus1 > 0
    else float("nan")
)

# ----- Write plain-text report -----
with open(file_report, "w", encoding="utf-8") as f:
    f.write("Overall statistics (all rows):\n")
    f.write(f"Number of answers: {n_answers}\n")
    f.write(f"Number of correct answers: {n_correct}\n")
    f.write(f"Percentage correct: {pct_correct:.2f}%\n\n")

    f.write('Statistics ignoring rows where Score is "-1":\n')
    f.write(f"Number of answers (Score != -1): {n_answers_no_minus1}\n")
    f.write(f"Number of correct answers (Score != -1): {n_correct_no_minus1}\n")
    f.write(f"Percentage correct (Score != -1): {pct_correct_no_minus1:.2f}%\n")
```